\documentclass[conference]{IEEEtran}
\makeatletter
\renewcommand{\@IEEEsectpunct}{\ \,} 
\makeatother
\usepackage[table,xcdraw]{xcolor}
\usepackage{url}
\usepackage{amsmath} 
\usepackage{graphicx} 
\usepackage{cite} 
\usepackage{color} 
\usepackage{soul} 
\usepackage[]{todonotes} 
\usepackage{bigstrut} 
\usepackage{listings} 
\usepackage{balance} 
\usepackage{algpseudocode} 
\usepackage{algorithm} 
\usepackage{algorithmicx} 
\usepackage{multicol}
\usepackage{setspace}
\usepackage{listings} 

\usepackage{amsthm} 
\usepackage{multirow} 
\usepackage{rotating} 
\usepackage{verbatim}
\usepackage{rotating} 
\usepackage{framed}
\usepackage{alltt}
\usepackage{graphicx}
\usepackage{lipsum}
\usepackage{color}
\usepackage{subfigure}
\usepackage{cite}
\usepackage{tikz}

\newcommand\numAttributes{19}

\newcommand\numCVEs{365}
\newcommand\numClassifiers{6}

\begin{document}

\title{Automated Characterization of Software Vulnerabilities}

 \author{ Danielle Gonzalez, Holly Hastings, Mehdi Mirakhorli \\  Rochester Institute of Technology, Rochester, NY  \\ 
 \{dng2551,hmh6182,mxmvse\}@rit.edu}
 
\maketitle
\begin{abstract}
Preventing vulnerability exploits is a critical software maintenance task, and software engineers often rely on Common Vulnerability and Exposure (CVEs) reports for information about vulnerable systems and libraries. These reports include descriptions, disclosure sources, and manually-populated vulnerability characteristics such as root cause from the NIST Vulnerability Description Ontology (VDO).  This information needs to be complete and accurate so stakeholders of affected products can prevent and react to exploits of the reported vulnerabilities. 
However, characterizing each report requires significant time and expertise which can lead to inaccurate or incomplete reports. This directly impacts stakeholders ability to quickly and correctly maintain their affected systems. 

In this study, we demonstrate that VDO characteristics can be automatically detected from the textual descriptions included in CVE reports. We evaluated the performance of 6 classification algorithms with a dataset of 365 vulnerability descriptions, each mapped to 1 of 19 characteristics from the VDO. This work demonstrates that it is feasible to train classification techniques to accurately characterize vulnerabilities from their descriptions. All 6 classifiers evaluated produced accurate results, and the Support Vector Machine classifier was the best-performing individual classifier. Automating the vulnerability characterization process is a step towards ensuring stakeholders have the necessary data to effectively maintain their systems. 
\end{abstract}



\IEEEpeerreviewmaketitle

\section{Introduction}
\label{sec:intro}
Preventing exploits of existing and emergent vulnerabilities is a critical software maintenance task. Whenever a new vulnerability or exploit affecting a software product is discovered, the software must be updated to remove the vulnerability and/or add appropriate mitigation technique. Stakeholders of a software product must have access to comprehensive and accurate data about vulnerabilities affecting their system to perform this maintenance. The standardized source for vulnerability data are Common Vulnerability and Exposure (CVE) reports, stored in the online and publicly available National Vulnerability Database~\cite{booth2013national} maintained by the National Institute of Standards and Technology (NIST). Listing 1 is an example of a vulnerability description from a NIST CSV report. 


\begin{framed}
	\vspace{-5pt}
    \scriptsize
\begin{spacing}{0.7}
\noindent\underline{\textbf{CVE ID}}: CVE-2017-6725 \\
	\noindent\underline{\textbf{Overview}}:
   A vulnerability in the web framework code of Cisco Prime Infrastructure could allow an unauthenticated, remote attacker to conduct a cross-site scripting (XSS) attack against a user of the web interface of an affected system. More Information: CSCuw65833 CSCuw65837. Known Affected Releases: 2.2(2). \\
	\noindent\underline{\textbf{References}}: \url{http://www.securityfocus.com/bid/99202}, \textbf{[...]} \\
	\normalsize
    \vspace{-8pt}
    \end{spacing}
\end{framed}
\vspace{-3pt}
\small
\noindent \textbf{Listing 1.} Vulnerability Description from the NVD
\normalsize\vspace{6pt}

Characterizing software vulnerabilities is a critical step in identifying the root cause of the vulnerability, understanding its consequences, attack mechanisms and appropriate mitigation techniques. A vulnerability characterization is an important property for making informed decisions to fix or mitigate the vulnerability. NIST has developed a standardized \textit{Vulnerability Description Ontology} (VDO)~\cite{booth2016draft} to characterize software vulnerabilities. The VDO defines the description attributes required to effectively provided actionable intelligence to the software developers aiming to fix/mitigate the vulnerability.

\begin{figure}
\centering
\includegraphics[width=\linewidth]{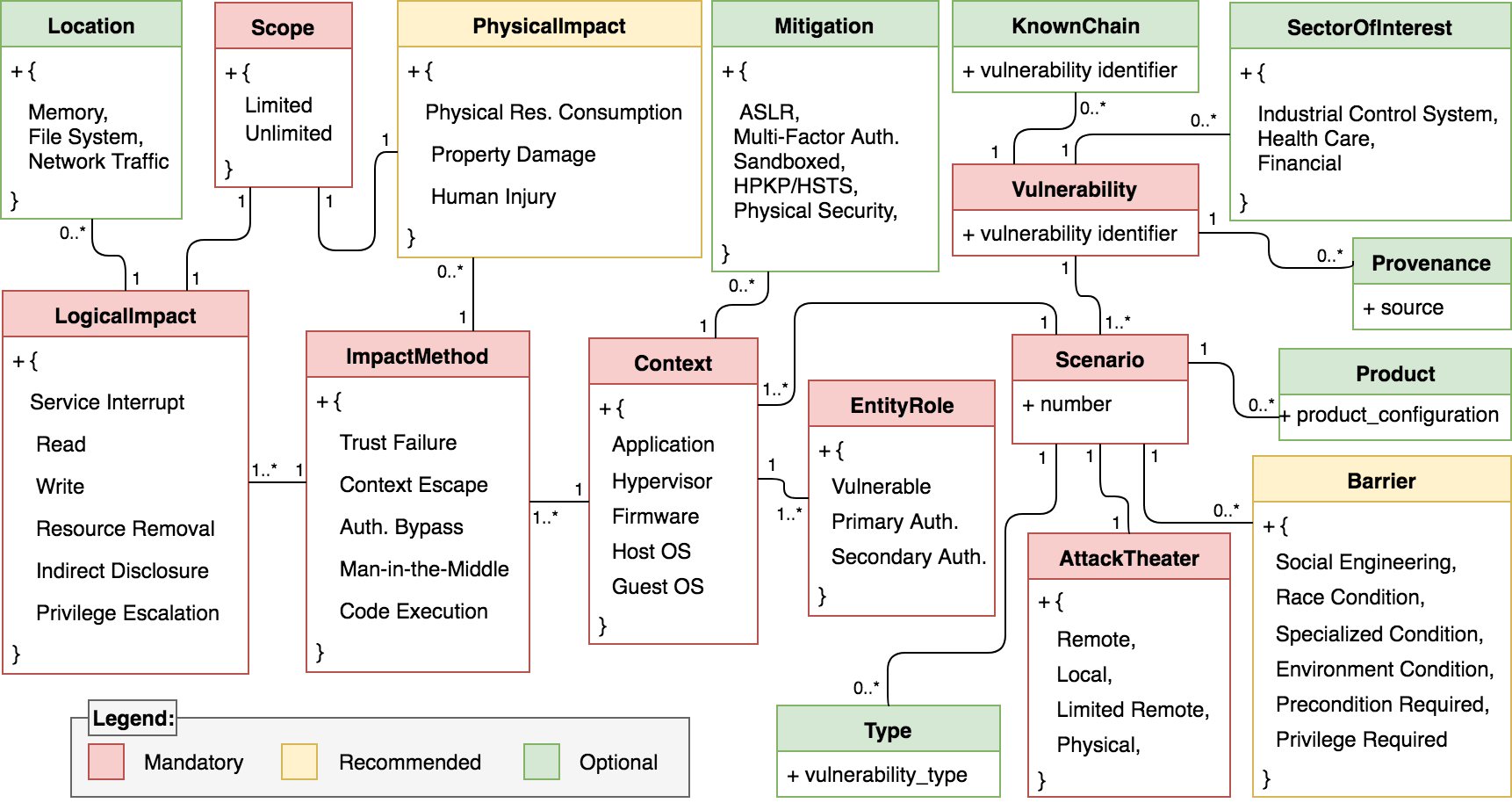}
\centering\caption{The NIST Vulnerability Description Ontology (VDO)}
\label{fig:vdoDiagram}
\end{figure}

Figure~\ref{fig:vdoDiagram} demonstrates various attributes of VDO ontology used to characterize software vulnerabilities. For example, the \textit{Impact Method} characterization category represents how a vulnerability can be exploited. The characterizations in the category reflect specific techniques an attacker can use to take advantage of a vulnerability: \textit{Authentication Bypass}, \textit{Trust Failure}, \textit{Context Escape}, \textit{Man-in-the-Middle Attack}, and \textit{Code Execution}. Other characterizations specify consequences, domain, location, and mitigations for vulnerability exploits. 

Stakeholders and affected users of the vulnerable product rely on characteristic data to determine how an exploit affects their systems and how to prevent exploitation. Unfortunately, identifying a vulnerability's characteristics as described in the VDO is a manual, labor-intensive, and asynchronous process. Proper characterization of a vulnerability requires reviewing its descriptions with sufficient security background and familiarity with the ontology. This intensive process has lead to problems with the quality of the vulnerability reports. Studies have found that NVD's vulnerability reports are often left incomplete or lack specific characteristics~\cite{Massacci:2010:RSV:1853919.1853925,ozment2007vulnerability}. 

In this study, we  conducted an experiment in which we used information retrieval, natural language processing, and supervised machine learning techniques to characterize vulnerabilities based on the descriptions present in every CVE report. We manually curated a dataset of \numCVEs{} vulnerability descriptions, each mapped to 1 of \numAttributes{} characteristics from the NIST Vulnerability Description Ontology (VDO). This data was run through several  stratified 10-fold cross validation experiments to train and evaluate the performance of \numClassifiers{} classification algorithms to determine which was most suited for the task. We initially found that the classifier with the best performance was a Majority Vote ensemble learner, which assigned the majority-chosen label from 1 instance each of the Na\"{i}ve Bayes, Decision Tree, Support Vector Machine (SVM), AdaBoost-SVM, and Random Forest classifiers. However, further statistical tests showed that it was not significantly better than an individual Support Vector Machine classifier, which requires less resources to train.  

We conclude from our findings that it is feasible to rely only on vulnerability description to characterize the CVE attributes. This automated approach can provide valuable information to the software developers so they can better understand a reported vulnerability and its characteristics.

The remainder of this paper is organized as follows. 
Our data collection and experiment methodology are explained in Section~\ref{sec:method}.
The results of our experiment are reported and discussed in Section~\ref{sec:results}.
We acknowledge related work in Section~\ref{sec:related}, and we conclude in Section~\ref{sec:conc}.
\section{Automated Characterization of Vulnerability Reports}
\label{sec:method}
Towards the goal of automating the vulnerability characterization process, we developed an approach that applies natural language processing (NLP), information retrieval (IR), and supervised machine learning techniques to train a classifier to analyze CVE reports and automatically infer their VDO characteristics. This is based on an assumption that the terms used in CVE descriptions offer insights into these characteristics of the vulnerability, and classifiers can learn such insight based on historical data. To test the feasibility of this approach, we trained multiple classification algorithms and compared their performance to determine the best performing classifier for this specific task.

In this section we describe our evaluation approach. First, we created a labeled dataset by manually curating relevant vulnerability descriptions for each of the \numAttributes{} characteristics 
Next, \numClassifiers{} classifiers 
were trained and tested using stratified 10-fold cross validation. 
The results were used to calculate a set of evaluation metrics, which were used to compare the classifiers' performance as reported in Section~\ref{sec:results}.


\subsection{Labeled Dataset Creation}
\label{sec:trainingData}
Using textual descriptions to characterize vulnerability reports is considered a multi-class text classification task~\cite{sokolova2009systematic}. In binary and multi-class classification, each item in the training dataset is manually assigned 1 class (label). Therefore, our first task was to curate a set of vulnerability descriptions for each of the \numAttributes{} characteristics.


We peer-reviewed numerous CVEs and their vulnerability descriptions in the National Vulnerability Database (NVD)~\cite{booth2013national} and selected CVE reports that had previously been characterized by the original developers. These CVE reports and their instantiated characteristics were peer reviewed by two  security experts from our research group. Descriptions were only kept if the two members of our team agreed with the labeling performed by the original developer. The end result was a manually curated training dataset of \numCVEs{} labeled descriptions. The number of CVE descriptions mapped to each VDO attribute ranged from 12 to 26, and the median was 19.  
To mitigate the risks associated with imbalanced data, Proper precautions were taken to ensure that the evaluations handled this situation appropriately (see Section~\ref{sec:classifierRun} below). Figure~\ref{fig:dataDist} shows the distribution of descriptions across the characteristics.  
\begin{figure}[ht]
\centering
\includegraphics[width=120px,height=120px]{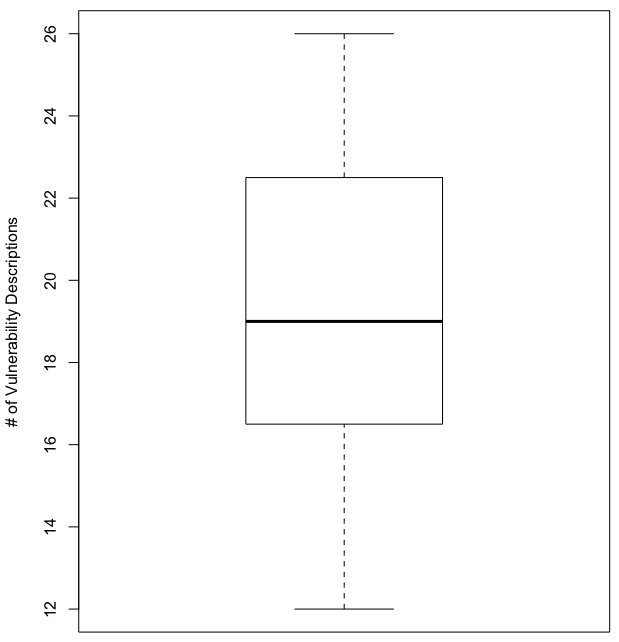}
\caption{Distribution of Vuln. Descriptions per Characteristic}
\vspace{4pt}
\label{fig:dataDist}
\end{figure}

\subsection{Training Data Preparation}
Before the training data can be provided as input to the classification algorithms, it must be \textit{pre-processed}. This is a necessary step in the natural language processing (NLP) data pipeline which cleans the text data and converts it to a numerical representation for use in classic machine learning algorithms. 
The vulnerability descriptions mined from the NVD's online vulnerability reports are considered ``free text'', meaning there were limited restrictions on what characters were permitted. This means it can be ``dirty" and contain non-textual characters which will affect the quality of the classifications if left in the input data. 

To prepare our text data for classification, we cleaned each vulnerability description. First, all characters were converted to lowercase and urls were removed. Next, the text was \textit{tokenized} by splitting sentences in the text by spaces into individual words. The set of tokens were then filtered to remove any non-word tokens (eg.!,\#,?). We also removed stop words (eg. ``the",``a"). Next, each word in a vulnerability description was \textit{stemmed} to its root. Then the text data was converted to a numerical representation known as a \textit{Term Frequency-Inverse Document Frequency (TF-IDF) Matrix}. Each row in the matrix is a vulnerability description, and each pre-processed word in the entire dataset is a column. Therefore, the TF-IDF weights in the matrix are calculated per-description for every word. These values are proportions of how often the word occurs within the description and how many descriptions it occurs in. 
This matrix is then provided as input to each classification algorithm. For our work, we used the StringToWordVec filter in the Weka tool~\cite{hall2009weka} to perform all these operations.

\subsection{Classification}
We compared the performance of \numClassifiers{} classification algorithms, using implementations from the Weka data mining tool. We ran preliminary experiments of all algorithms available in the tool and empirically selected 3 ``individual" classifiers (Na\"{i}ve Bayes, Decision Tree, Support Vector Machine) and 3 ``ensemble" classifiers (Random Forest, AdaBoost-SVM, Majority Vote). 

\subsection{Stratified 10-Fold Cross-Validation}
\label{sec:classifierRun}
Cross validation is a standard technique for training a classifier and evaluate its performance for unseen data. To mitigate risks related to imbalanced data~\cite{Forman:2010:ACS:1882471.1882479}, we applied \textit{stratified} 10-fold cross validation which splits the labeled data into 10 sets, representing each class proportionally in each set. 
To tune the hyper-parameters for each of the \numClassifiers{} classifiers, we ran the 10-fold experiments using multiple configurations. The best performing configurations were used in our evaluation. These configurations are provided in Table~\ref{tab:wekaSchemes} in the Weka~\cite{hall2009weka} ``scheme" format used to run the experiments.

\begin{table*}
\centering
\caption{Classification Algorithm Hyper-Parameter Configuration Schemes Used in Weka Tool}
\begin{tabular}{|l|p{15.3cm}|}
\hline
\textbf{\small{Classifier}} & \textbf{\small{Weka Schema (Hyper-parameters)}}                                                                                                     \\ \hline
\small{Random Forest}       & \small{weka.classifiers.trees.RandomForest,-P 100 -O -I 320 -num-slots 1 -K 1 -M 1.0 -V 0.001 -S 123 -B}                                                                                                                                                                                                             \\ \hline
\small{Decision Tree}       & \small{weka.classifiers.trees.J48 -C 0.4 -M 0 }                                                                                             \\ \hline
\small{SVM}                 & \small{weka.classifiers.functions.SMO -C 0.5 -L 0.001 -P 1.0E-12 -N 0 -V 10 -W 123 -K ``weka.classifiers.functions.supportVector.PolyKernel -E 1.0 -C 250007" -calibrator ``weka.classifiers.functions.Logistic -R 1.0E-8 -M -1 -num-decimal-places 4"}                                                                                                                                             \\ \hline
\small{AdaBoost-SVM}            & \small{weka.classifiers.meta.AdaBoostM1,-P 100 -S 123 -I 100 -W weka.classifiers.functions.SMO -- -C 0.5 -L 0.001 -P,1.0E-12 -N 0 -V 10 -W 123 -K,``weka.classifiers.functions.supportVector.PolyKernel -E 1.0 -C,250007" -calibrator ``weka.classifiers.functions.Logistic -R 1.0E-8,-M -1 -num-decimal-places 4"}                                                                                       \\ \hline
\small{MajorityVote}        & \small{weka.classifiers.meta.Vote,-S 123 -B ``weka.classifiers.bayes.NaiveBayes " -B,``weka.classifiers.functions.SMO -C 0.5 -L 0.001 -P 1.0E-12 -N 0 -V 10 -W,123 -K \textbackslash{}``weka.classifiers.functions.supportVector.PolyKernel -E 1.0 -C,250007\textbackslash{}" -calibrator \textbackslash{}``weka.classifiers.functions.Logistic -R,1.0E-8 -M -1 -num-decimal-places 4\textbackslash{}"" -B,``weka.classifiers.trees.J48 -C 0.4 -M 0" -B,``weka.classifiers.trees.RandomForest -P 100 -O -I 320 -num-slots 1 -K 1,-M 1.0 -V 0.001 -S 123" -B ``weka.classifiers.meta.AdaBoostM1 -P 100,-S 123 -I 100 -W weka.classifiers.functions.SMO -- -C 0.5 -L 0.001 -P 1.0E-12,-N 0 -V -1 -W 123 -K,\textbackslash{}``weka.classifiers.functions.supportVector.PolyKernel -E 1.0 -C,250007\textbackslash{}" -calibrator \textbackslash{}``weka.classifiers.functions.Logistic -R,1.0E-8 -M -1 -num-decimal-places 4\textbackslash{}"" -R MAJ} \\ \hline
\end{tabular}
\label{tab:wekaSchemes}
\end{table*}
\section{Results}
\label{sec:results}
After the stratified 10-fold cross validation was conducted for all classifiers, 
we calculated 6 standard evaluation metrics from Weka's output to compare their performance. 

The baseline metrics for each classifier are shown in Table~\ref{tab:baseline}. The Majority Vote ensemble classifier had the highest accuracy, or total percentage of correctly classified data, at 74.52\%. Majority Vote also had the highest Kappa statistic of 0.73. SVM and AdaBoost-SVM tied for 2nd best accuracy (72.88\%) and Kappa (0.71).
\begin{table}[ht]
\centering
\caption{Baseline Metrics For Classifiers}
\label{tab:baseline}
\begin{tabular}{|l|c|c|}
\hline
\textbf{Classifier}  & \textbf{Accuracy} & \textbf{Kappa statistic} \\ \hline
Naïve Bayes   & 66.85\%           & 0.65                  \\ \hline
Decision Tree & 64.93\%           & 0.63                   \\ \hline
SMO-SVM       & 72.88\%           & 0.71                   \\ \hline
Random Forest & 71.51\%           & 0.70                  \\ \hline
AdaBoost-SVM  & 72.88\%           & 0.71                   \\ \hline
\textbf{Majority Vote}  & 74.52\%           & 0.73                   \\ \hline
\end{tabular}%

\end{table}

Per-characteristic information retrieval metrics (precision, recall, and F-measure) were computed for each of the 6 classifiers as shown in Table~\ref{tab:results}. The best F-measure for each characteristic is highlighted (ties are all counted as best) and the number of highlights per classifier was used to calculate what we call the \textit{Ratio of Best Performance} (RBF) for each classifier. 
This a ratio of the number of characteristics that the classifier ``won" (highest F1) to the total number of characteristics evaluated (\numAttributes{}). Using this metric, the Majority Voting ensemble was the best performing classifier,   for 8 of the characteristics ($RBF = 0.42$).

\begin{table*}
\caption{Per-characteristic Results of Stratified 10-Fold Cross Validation}
\label{tab:results}
\resizebox{\textwidth}{!}{%
\begin{tabular}{cccccccccccccccccccccccc}
\cline{1-12} \cline{14-24}
\multicolumn{12}{c}{\textbf{Individual Classifiers}}                                                                                                                                        &  & \multicolumn{11}{c}{\textbf{Ensemble Classifiers}}                                                                                                                           \\ \cline{1-12} \cline{14-24} 
                          & \multicolumn{3}{c}{\textbf{Na\"{i}ve Bayes}}          &  & \multicolumn{3}{c}{\textbf{Decision Trees}}       &  & \multicolumn{3}{c}{\textbf{SVM}}                  &  & \multicolumn{3}{c}{\textbf{Random Forest}}        &  & \multicolumn{3}{c}{\textbf{AdaBoost-SVM}}         &  & \multicolumn{3}{c}{\textbf{Majority Vote}}                     \\ \cline{2-4} \cline{6-8} \cline{10-12} \cline{14-16} \cline{18-20} \cline{22-24} 
\textbf{characteristic} & Precision & Recall & F1                           &  & Precision & Recall & F1                           &  & Precision & Recall & F1                           &  & Precision & Recall & F1                           &  & Precision & Recall & F1                           &  & Precision      & Recall     & F1                               \\ \cline{1-4} \cline{6-8} \cline{10-12} \cline{14-16} \cline{18-20} \cline{22-24} 
ASLR                      & 1.00      & 0.80   & 0.89                         &  & 0.75      & 0.75   & 0.75                         &  & 1.00      & 0.75   & 0.86                         &  & 1.00      & 0.80   & 0.89                         &  & 1.00      & 0.75   & 0.86                         &  & 1.00           & 0.85       & \cellcolor[HTML]{C0C0C0}0.92     \\
Context Escape            & 0.76      & 0.96   & 0.85                         &  & 0.77      & 0.65   & 0.71                         &  & 0.85      & 0.85   & 0.85                         &  & 0.77      & 1.00   & \cellcolor[HTML]{C0C0C0}0.87 &  & 0.85      & 0.85   & 0.85                         &  & 0.83           & 0.92       & \cellcolor[HTML]{C0C0C0}0.87     \\
File System               & 0.83      & 0.42   & \cellcolor[HTML]{C0C0C0}0.56 &  & 0.30      & 0.25   & 0.27                         &  & 0.80      & 0.33   & 0.47                         &  & 0.43      & 0.25   & 0.32                         &  & 0.80      & 0.33   & 0.47                         &  & 0.80           & 0.33       & 0.47                             \\
HPHK                      & 0.64      & 0.75   & 0.69                         &  & 0.88      & 0.58   & 0.70                         &  & 0.89      & 0.67   & 0.76                         &  & 0.89      & 0.67   & 0.76                         &  & 0.89      & 0.67   & 0.76                         &  & 0.90           & 0.75       & \cellcolor[HTML]{C0C0C0}0.82     \\
HSTS                      & 0.80      & 0.73   & 0.76                         &  & 0.76      & 0.86   & \cellcolor[HTML]{C0C0C0}0.81 &  & 0.59      & 0.77   & 0.67                         &  & 0.56      & 0.86   & 0.68                         &  & 0.59      & 0.77   & 0.67                         &  & 0.66           & 0.86       & 0.75                             \\
Indirect Disclosure       & 0.69      & 0.92   & 0.79                         &  & 1.00      & 0.63   & 0.77                         &  & 0.96      & 0.88   & \cellcolor[HTML]{C0C0C0}0.91 &  & 0.88      & 0.92   & 0.90                         &  & 0.96      & 0.88   & \cellcolor[HTML]{C0C0C0}0.91 &  & 0.91           & 0.88       & 0.89                             \\
Limited                   & 0.33      & 0.29   & 0.31                         &  & 0.36      & 0.29   & 0.32                         &  & 0.55      & 0.43   & \cellcolor[HTML]{C0C0C0}0.48 &  & 0.50      & 0.29   & 0.36                         &  & 0.55      & 0.43   & \cellcolor[HTML]{C0C0C0}0.48 &  & 0.55           & 0.43       & \cellcolor[HTML]{C0C0C0}0.48     \\
Man-in-the-Middle         & 0.75      & 0.71   & 0.73                         &  & 0.93      & 0.82   & \cellcolor[HTML]{C0C0C0}0.88 &  & 0.77      & 0.77   & 0.77                         &  & 0.77      & 0.77   & 0.77                         &  & 0.77      & 0.77   & 0.77                         &  & 0.81           & 0.77       & 0.79                             \\
Memory                    & 0.56      & 0.41   & 0.47                         &  & 0.58      & 0.68   & 0.63                         &  & 0.64      & 0.64   & \cellcolor[HTML]{C0C0C0}0.64 &  & 0.48      & 0.55   & 0.51                         &  & 0.64      & 0.64   & \cellcolor[HTML]{C0C0C0}0.64 &  & 0.64           & 0.64       & \cellcolor[HTML]{C0C0C0}0.64     \\
Multi-Factor Auth         & 0.69      & 0.64   & 0.67                         &  & 1.00      & 0.79   & \cellcolor[HTML]{C0C0C0}0.88 &  & 0.91      & 0.71   & 0.80                         &  & 0.82      & 0.64   & 0.72                         &  & 0.91      & 0.71   & 0.80                         &  & 0.91           & 0.71       & 0.80                             \\
Network Traffic           & 0.53      & 0.63   & 0.57                         &  & 0.53      & 0.56   & 0.55                         &  & 0.89      & 0.50   & \cellcolor[HTML]{C0C0C0}0.64 &  & 0.88      & 0.44   & 0.58                         &  & 0.89      & 0.50   & \cellcolor[HTML]{C0C0C0}0.64 &  & 0.89           & 0.50       & \cellcolor[HTML]{C0C0C0}0.64     \\
Physical Security         & 0.71      & 0.56   & 0.63                         &  & 0.82      & 0.78   & 0.80                         &  & 0.94      & 0.83   & \cellcolor[HTML]{C0C0C0}0.88 &  & 1.00      & 0.67   & 0.80                         &  & 0.94      & 0.83   & \cellcolor[HTML]{C0C0C0}0.88 &  & 0.94           & 0.83       & \cellcolor[HTML]{C0C0C0}0.88     \\
Privilege Escalation      & 0.91      & 0.87   & \cellcolor[HTML]{C0C0C0}0.89 &  & 0.86      & 0.83   & 0.84                         &  & 0.95      & 0.83   & 0.88                         &  & 0.86      & 0.83   & 0.84                         &  & 0.95      & 0.83   & 0.88                         &  & 0.95           & 0.83       & 0.88                             \\
Read                      & 0.77      & 0.71   & \cellcolor[HTML]{C0C0C0}0.74 &  & 0.47      & 0.63   & 0.54                         &  & 0.64      & 0.75   & 0.69                         &  & 0.64      & 0.75   & 0.69                         &  & 0.62      & 0.75   & 0.68                         &  & 0.69           & 0.75       & 0.72                             \\
Sandboxed                 & 0.61      & 0.58   & 0.60                         &  & 0.88      & 0.79   & \cellcolor[HTML]{C0C0C0}0.83 &  & 0.79      & 0.79   & 0.79                         &  & 0.82      & 0.74   & 0.78                         &  & 0.83      & 0.79   & 0.81                         &  & 0.79           & 0.79       & 0.79                             \\
Service Interrupt         & 0.68      & 0.77   & 0.72                         &  & 0.93      & 0.82   & 0.88                         &  & 0.94      & 0.88   & \cellcolor[HTML]{C0C0C0}0.91 &  & 0.76      & 0.94   & 0.84                         &  & 0.94      & 0.88   & \cellcolor[HTML]{C0C0C0}0.91 &  & 0.89           & 0.94       & \cellcolor[HTML]{C0C0C0}0.91     \\
Trust Failure             & 0.57      & 0.77   & 0.66                         &  & 0.33      & 0.62   & 0.43                         &  & 0.74      & 0.89   & \cellcolor[HTML]{C0C0C0}0.81 &  & 0.76      & 0.85   & 0.80                         &  & 0.74      & 0.89   & \cellcolor[HTML]{C0C0C0}0.81 &  & 0.70           & 0.89       & 0.78                             \\
Unlimited                 & 0.71      & 0.55   & 0.62                         &  & 0.61      & 0.64   & 0.62                         &  & 0.52      & 0.77   & 0.62                         &  & 0.64      & 0.73   & \cellcolor[HTML]{C0C0C0}0.68 &  & 0.52      & 0.77   & 0.62                         &  & 0.57           & 0.73       & 0.64                             \\
Write                     & 0.19      & 0.24   & 0.21                         &  & 0.09      & 0.06   & 0.07                         &  & 0.23      & 0.35   & 0.28                         &  & 0.29      & 0.29   & \cellcolor[HTML]{C0C0C0}0.29 &  & 0.23      & 0.35   & 0.28                         &  & 0.22           & 0.29       & 0.25                             \\ \cline{1-4} \cline{6-8} \cline{10-12} \cline{14-16} \cline{18-20} \cline{22-24} 
                          & \multicolumn{3}{c}{\textbf{RBF: 0.16}}            &  & \multicolumn{3}{c}{\textbf{RBF: 0.21}}            &  & \multicolumn{3}{c}{\textbf{RBF: 0.37}}            &  & \multicolumn{3}{c}{\textbf{RBF: 0.16}}            &  & \multicolumn{3}{c}{\textbf{RBF: 0.37}}            &  & \multicolumn{3}{c}{\cellcolor[HTML]{C0C0C0}\textbf{RBF: 0.42}} \\ \cline{2-4} \cline{6-8} \cline{10-12} \cline{14-16} \cline{18-20} \cline{22-24} 
\end{tabular}%
}
\vspace{-10pt}
\end{table*}

\subsection{Analysis of Results}
The baseline metrics in Table~\ref{tab:baseline} indicate that all classifiers had an accuracy higher than 50\%, and outperformed random assignment. While these metrics can be used to measure the performance of each classifier individually, they are not suitable for \textit{comparing} classifiers. However, these findings indicate that text classification algorithms are well-suited for the task of automatically characterize vulnerability reports using their description. 

The best performance metrics in Table~\ref{tab:results} were used instead to compare classifiers, and these results show that each classification algorithm performed best for at least 3 of the characteristic classes. The worst performing classifiers in this respect were Na\"{i}ve Bayes and Random Forest, each with $RBF=0.16$. The individual Support Vector Machine (SVM) and the ensemble AdaBoost-SVM had identical F-measures, tying for second-best classifier by performing best for 7 of 19 classes. These 2 classifiers also had the same accuracy and kappa statistic, which indicates that applying Adaptive Boosting did not improve the SVM classifier at all. 

To confirm our RBP-based findings, we also conducted statistical significance tests. First, we conducted a Friedman test~\cite{sokolova2009systematic} which confirmed a statistically significant ($p<0.05$) performance difference between \textit{any} of the classifiers, as shown in Table~\ref{tab:friedman}. We also performed a post-hoc Conover test~\cite{conover1999practical} to investigate if differences in performance between individual classifiers was statistically significant. The p-values from this analysis are listed in Table~\ref{tab:conover}. This data indicates that the SVM classifier's performance was significantly ($P < 0.05$) different from the Random Forest and Decision tree classifiers but not the Majority Vote or AdaBoost-SVM. 
\begin{table}[ht]
\centering
\caption{Testing for Significance of Classifier Performance}
\begin{tabular}{|l|l|}
\hline
\textbf{\# of Classes} & \numAttributes{}       \\ \hline
\textbf{Friedman's Chi-Squared}   & 19.38312 \\ \hline
\textbf{df}            & 5        \\ \hline
\textbf{p-value}       & 0.002    \\ \hline
\end{tabular}%
\label{tab:friedman}
\end{table}
\begin{table}
\caption{P-Values from Conover Post-Hoc Test}
\resizebox{\linewidth}{!}{%
\begin{tabular}{|l|l|l|l|l|}
\hline
 & \textbf{Decision Tree} & \textbf{SVM} & \textbf{Random Forest} & \textbf{Adaboost-SVM} \\ \hline
\textbf{SVM} & 0.000033072 & NA & NA & NA \\ \hline
\textbf{Random Forest} & 0.552778600 & 0.02808133 & NA & NA \\ \hline
\textbf{Adaboost-SVM} & 0.000033072 & 1.00 & 0.02808133 & NA \\ \hline
\textbf{Majority Vote} & 0.000000002 & 0.18663475 & 0.00000546 & 0.1866348 \\ \hline
\end{tabular}%
}
\label{tab:conover}
\end{table}

The statistical tests confirm that while the Majority Vote classifier had the highest accuracy and Ratio of Best Performance, it did not significantly outperform the individual Support Vector Machine (SVM). Therefore, there is no motivation to spend more resources training the ensemble Majority Vote and instead a Support Vector Machine is suited for this classification task.

\begin{framed}
\vspace{-6pt}
\noindent
\centering
\textbf{Classifier Evaluation Key Findings:}
\begin{itemize}
    \item Classification learners can be trained to accurately characterize vulnerability reports using their free-text descriptions.
    \item The classifier with the highest accuracy and Ratio of Best Performance in our evaluation was the Majority Vote ensemble learner, with a Ratio of Best Performance of 0.42 (performing best for 8 of \numAttributes{} characteristics) and overall accuracy of 74.52\%. 
    \item The Support Vector Machine (SVM) and AdaBoost-SVM classifiers tied for 2nd-best with a Ratio of Best Performance of 0.37. 
    \item Majority Vote did not \textit{significantly} outperform these classifiers, and the AdaBoost-SVM ensemble did not \textit{significantly} outperform the individual Support Vector Machine. 
    \item Considering these evaluations and time-to-train,an individual \textbf{Support Vector Machine is concluded to be the best-performing classifier for this task}.
\end{itemize}
\vspace{-6pt}
\end{framed}
\section{Related Work}
\label{sec:related}

\subsection{Analysis of Vulnerability Report Data}
Several studies have evaluated the \textit{quality} of vulnerability reports which motivated our work to improve them. An early and extensive study by Ozment~\cite{ozment2007vulnerability} noted the inconsistent terminology used at the time for characterizing vulnerabilities and also highlighted the inconsistent report quality. Massacci and Nguyen~\cite{Massacci:2010:RSV:1853919.1853925} compared characterizations included in reports from 14 vulnerability databases, maintained by specific and multi-software vendors (eg. Bugzilla and NVD). They corroborated the earlier findings, noting NVD lacked or had incomplete data for many temporal and code-related characterizations. Ladd~\cite{RecordedFutureCNNVD} compared the report generation processes of the NVD and its Chinese equivalent the CNNVD, concluding that the asynchronous reporting processes of the NVD cause the US falls behind on vulnerability management.

\subsection{Characterizing Vulnerabilities}
Other relevant works aimed to develop or identify of vulnerability characterizations, but do not use the NIST Vulnerability Description Ontology, which was published in late 2016~\cite{booth2016draft}. Tierney~\cite{tierney2005knowledge} used similar data mining and machine learning techniques and CVE report data, including a precursor of NIST's CVSS severity scores~\cite{4042667}, to learn vulnerability patterns and co-ocurring vulnerabilities, and to classify vulnerabilities by severity. Zhang et al.~\cite{zhang2015predicting} extracted temporal, affected version, and severity score data from CVE reports and evaluated 6 regression and classification algorithms on the task of predicting a system's time to next vulnerability with limited success, which they attributed to the poor quality of the NVD data. Edkrantz and Said~\cite{edkrantz2015predicting} extracted \textit{n-grams} from CVE report summaries along with other CVE data such as the severity score to evaluate the performance of 5 classification algorithms on the task of predicting the likelihood that a vulnerability would be exploited. SVM classifiers also performed best for their task. Santos et. al~\cite{santos2017catalog} curated a catalog of \textit{architectural} weaknesses leveraging root cause data from vulnerability reports. Gonzalez et. al~\cite{8703918} also analyzed root cause data from vulnerability reports to identify weaknesses common in the domain of Industrial Control Systems.

Most similar to our work is that of Joshi et al.~\cite{6693525}, who also characterized vulnerabilities using the textual descriptions from the CVE reports. While this work shares our goal, the approaches are very different. Instead of classifiers, they trained a custom version of Stanford CoreNLP's Named Entity Recognition implementation to identify and distinguish specific people, places, and things. They defined 10 characterization classes based on existing security literature, and also used terminology from a different ontology~\cite{more2012knowledge}.

\section{Conclusions \& Next Steps}
\label{sec:conc}
This study was motivated by the existing challenges is maintaining vulnerability reports. We evaluated an approach towards automating the process of identifying characteristics for vulnerabilities using their descriptions. We curated \numCVEs{} vulnerability descriptions from the National Vulnerability Database (NVD), each mapped to 1 of \numAttributes{} characteristics from the NIST Vulnerability Description Ontology. We used this data train and evaluate the performance of \numClassifiers{} \textit{classification} algorithms using stratified 10-fold cross validation. Standard information retrieval and machine learning metrics were used to identify the classification algorithm which performed best. 

We found that all of the classifiers were able to accurately identify relevant vulnerability characteristics from the NIST Vulnerability Description Ontology (VDO). The best performing classifier for this task based on accuracy and the Ratio of Best Performance metric was a Majority Vote ensemble learner, which assigns the majority-chosen label from 1 instance each of the Na\"{i}ve Bayes, Decision Tree, Support Vector Machine (SVM), AdaBoost-SVM, and Random Forest classifiers. However, further statistical tests showed that it was not significantly better than an individual Support Vector Machine classifier, which requires less resources to train. This is an encouraging finding towards measuring the suitability of automated text classification algorithms for this task. With a learner that does not require extensive resources to train and test, vulnerability data can be easily updated and expanded. 

With this work, we have demonstrated that the vulnerability management process can be improved with this approach thereby providing actionable information to programmers to effectively perform software engineering tasks. Future research will focus on expanding our dataset to include more descriptions and support the remaining characteristics, and conducting further classifier evaluations to increase the accuracy and scope of the automated approach. We also aim to utilize the CVE characteristics to generate intelligence about each vulnerability and consequently recommend appropriate mitigation strategies to software maintainers.


\vspace{-5pt}
\section*{Acknowledgments}
This work was partially funded by the US National Science Foundation under grant number CNS-1816845 and IIP-0968959 under funding from the S2ERC I/UCRC program and US Department of Homeland Security.
\bibliographystyle{abbrv}
\bibliography{Papers}

\end{document}